\documentclass[conference]{IEEEtran}
\IEEEoverridecommandlockouts

\usepackage{cite}
\usepackage{amsmath,amssymb,amsfonts}
\usepackage{algorithmic}
\usepackage{graphicx}
\usepackage{textcomp}
\usepackage{xcolor}
\usepackage{pdfpages}
\usepackage{subfloat}
\usepackage{subfigure}
\usepackage{svg}
\usepackage[T1]{fontenc}
\usepackage{xcolor}
\def\BibTeX{{\rm B\kern-.05em{\sc i\kern-.025em b}\kern-.08em
    T\kern-.1667em\lower.7ex\hbox{E}\kern-.125emX}}
\DeclareRobustCommand*{\IEEEauthorrefmark}[1]{
\raisebox{0pt}[0pt][0pt]{\textsuperscript{\footnotesize\ensuremath{#1}}}}
\begin{document}

\title{Collaborative Computing Strategy Based SINS Prediction for Emergency UAVs Network}

\author{\IEEEauthorblockN{Bing Li\IEEEauthorrefmark{1}, Haoming Guo\IEEEauthorrefmark{2}, Zhiyuan Ren\IEEEauthorrefmark{*,1}, Wenchi Cheng\IEEEauthorrefmark{1}, Jialin Hu\IEEEauthorrefmark{1}, Xinke Jian\IEEEauthorrefmark{1}}
\IEEEauthorblockA{\IEEEauthorrefmark{1}{State Key Laboratory of Integrated Services Networks,}
{Xidian University,}
Xi'an, China \\
\IEEEauthorrefmark{2}{Institute of Software Chinese Academy of Sciences,}
Beijing, China\\
Email:\{23011210593@stu.xidian.edu.cn, haoming@iscas.ac.cn, zyren@xidian.edu.cn,\\
wccheng@xidian.edu.cn, 22011210534@stu.xidian.edu.cn, xkjian@stu.xidian.edu.cn\}}
}

\maketitle

\begin{abstract}
    In emergency scenarios, the dynamic and harsh conditions necessitate timely trajectory adjustments for drones, leading to highly dynamic network topologies and potential task failures. To address these challenges, a collaborative computing strategy based strapdown inertial navigation system (SINS) prediction for emergency UAVs network (EUN) is proposed, where a two-step weighted time expanded graph (WTEG) is constructed to deal with dynamic network topology changes. Furthermore, the task scheduling is formulated as a Directed Acyclic Graph (DAG) to WTEG mapping problem to achieve collaborative computing while transmitting among UAVs. Finally, the binary particle swarm optimization (BPSO) algorithm is employed to choose the mapping strategy that minimizes end-to-end processing latency. The simulation results validate that the collaborative computing strategy significantly outperforms both cloud and local computing in terms of latency. Moreover, the task success rate using SINS is substantially improved compared to approaches without prior prediction.
\end{abstract}


\section{\bfseries INTRODUCTION}
In disaster scenarios, unmanned aerial vehicles (UAVs) have been widely applied in various fields owing to their high flexibility, rapid deployment capabilities, and cost-effectiveness \cite{b1}. To address the "information island" phenomenon caused by damaged infrastructure, UAVs can collect data from affected areas and then provide critical information for rescuers to save more lives. However, the collected raw data are often voluminous, necessitating further processing to extract valuable insights. A single UAV is often constrained by limited computational resources, making it unsuitable for complex data processing tasks \cite{b2}, while adopting a cloud server will bring high transmission latency \cite{b3, b4}. Therefore, constructing a UAV swarm system with collaborative edge computing capabilities can effectively enhance the overall system computing efficiency.

\let\thefootnote\relax\footnotetext{This work was supported in part by the National Key R$\&$D Program of China (2023YFC3011502)}
\let\thefootnote\relax\footnotetext{*Corresponding author: Zhiyuan Ren}

In the field of UAV cooperative computation offloading, current researches often lack dynamic adaptability. Existing algorithms mostly relied on an assumption of a fixed network topology \cite{b5, b6}, or pre-determined drone flight trajectories \cite{b7, b8}. For instance, Guo et al. \cite{b5} proposed an SDN-enhanced cooperative MUEAC system, where the UAVs keep hovering. Mohammad \cite{b6} designed a drone-fog-cloud collaborative strategy assuming a static network topology. Cheng et al. \cite{b7} proposed an air-space-ground integrated network architecture, where the UAVs were assigned fixed flying trajectories. Liu et al. \cite{b8} designed a drone transmission model in D2D communication, where the UAVs' trajectories remained invariable.

Given the highly dispersed user distribution in disaster areas, drones need to adjust their trajectories to provide continuous communication services for ground users. Moreover, the dynamic topology brings a significant challenge for the credibility of the task scheduling result, as the actual latency may be much higher than the predicted latency, potentially leading to task failures. Therefore, we propose a collaborative computing strategy for Emergency UAV Networks (EUN) based SINS prediction to enable simultaneous computation and transmission. The major contributions of our work are as follows:
\begin{itemize}
\item To overcome the topological uncertainty, we propose a spatio-temporal joint prediction model based on SINS, and develop the WTEG to effectively model the dynamic network topology.
\item To satisfy the latency requirements of tasks, we formulate the task scheduling problem as a DAG-WTEG mapping model. The BPSO algorithm is then employed to search for the optimal scheduling solution that minimizes end-to-end latency.
\end{itemize}

The rest of this paper is organized as follows. The system model is proposed in Section \uppercase\expandafter{\romannumeral2}. The details of BPSO are given in Section \uppercase\expandafter{\romannumeral3}. The simulation results and analysis are presented in Section \uppercase\expandafter{\romannumeral4}. Finally, Section \uppercase\expandafter{\romannumeral5} concludes the paper.

\section{\bfseries SYSTEM MODEL AND PROBLEM FORMULATION}

The system architecture of the air-ground network in disaster scenarios is depicted in Fig. 1. The system adopts a three-layer architecture design. The central control layer is composed of UAVs with high computing capability, which allocate subtasks to the UAVs in the subsequent layer according to the preset scheduling results. The airborne computing layer consists of UAVs with edge computing ability, which continuously adjust their trajectories to provide high-quality communication services for ground users. Finally, the terminal access layer consists of user devices in the disaster area, accessing the aerial computing layer via the air-ground link. Furthermore, the task is modeled as the DAG and orchestrated along transmission routes through the WTEG, enabling collaborative computing while transmitting among UAVs.
\vspace{-10pt}
\begin{figure}[htbp]
\setlength{\abovecaptionskip}{-2pt}
\centerline{\includegraphics[scale=0.5]{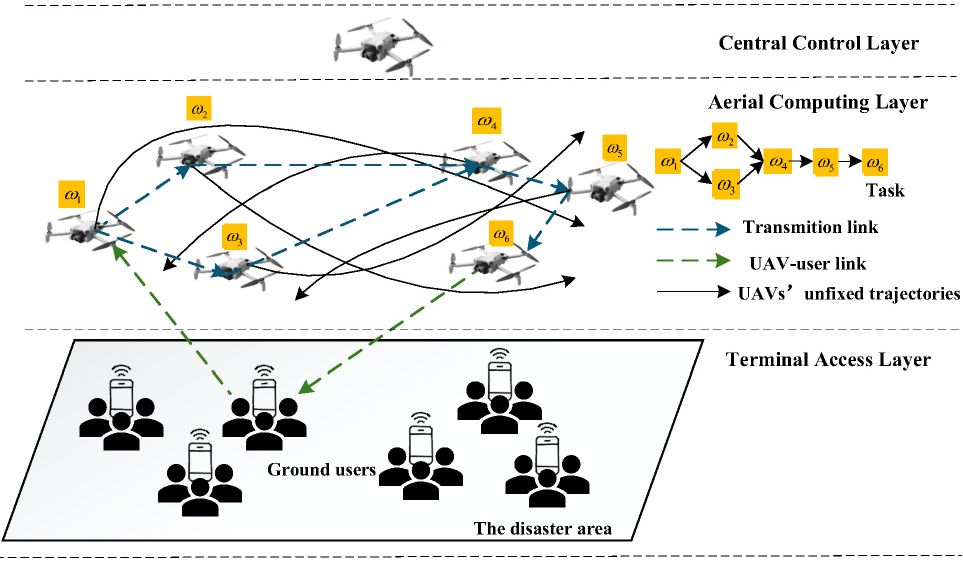}}
\caption{The system architecture of the Air-Ground Network}
\label{fig}
\end{figure}
\vspace{-10pt}

\subsection{Network Model}\label{AA}

In emergency rescue scenarios, the UAV clusters' trajectories are highly dynamic, leading to frequent changes in the network topology. The inherent dynamic nature of the network topology may result in service interruptions and task execution failures, particularly when using traditional service mapping approaches during time slot switches. Therefore, we first employ SINS to predict the drones' future positions with high prediction accuracy over short durations to construct a network topology prediction model. And then the Time Extended Graph (TEG) model is adopted to overcome the dynamic changes of EUN, which has been widely used in several fields \cite{b9, b10}. Finally, in order to further calculate the task processing latency, a two-step WTEG model is constructed by adding uniform delay weight parameters for the edges of TEG.

1) Mobility Prediction Model: Assuming that the network operation period is T, divided into n time slots. The network topology is assumed to be fixed within each time slot. The SINS, which comprises accelerometers and gyroscopes, offers a low-cost and highly accurate method for short-term position prediction \cite{b11}.

As illustrated in Fig. 2, the SINS mechanization process involves two parallel integration paths. First, the error-compensated angular velocity from the gyroscope is integrated to update the attitude matrix. This matrix then transforms the compensated acceleration from the carrier coordinate system (b) to the navigation coordinate system (n), which is subsequently integrated to update velocity and position. Furthermore, the subsequent velocity and position data facilitate the calculation of the platform's rotation rate, which is essential for the attitude matrix update.
The sensors sample twice at equal intervals during the time period $[t_{m-1},t_m]$, yielding angular increments $\Delta{\theta_{m1}}$ and $\Delta{\theta_{m2}}$, and velocity increments $\Delta{v_{m1}}$ and $\Delta{v_{m2}}$ in the carrier coordinate system(b). The basic equations for attitude, velocity, and position updates are as follows:

\vspace{-10pt}
\begin{equation}
\resizebox{0.9\hsize}{!}{$
C_{b}^{n}=
\begin{bmatrix}
cos{\gamma}cos{\varphi}+sin{\gamma}sin{\phi}sin{\theta} & sin{\varphi}cos{\theta} & sin{\gamma}cos{\varphi}-cos{\gamma}sin{\varphi}sin{\theta}\\
-cos{\gamma}sin{\varphi}+sin{\gamma}cos{\phi}sin{\theta} & cos{\varphi}cos{\theta} & -sin{\gamma}sin{\varphi}-cos{\gamma}cos{\varphi}sin{\theta}\\
-sin{\gamma}cos{\theta} & sin{\theta} & cos{\gamma}cos{\theta}
\end{bmatrix}
$}
\end{equation}
\vspace{-10pt}

\vspace{-18pt}
\begin{gather}
C_{b(m)}^{n(m)}=C_{n(m-1)}^{n(m)}C_{b(m-1)}^{n(m-1)}C_{b(m)}^{b(m-1)}\label{Eq.2}
\\
v_m^{n(m)}=v_{m-1}^{n(m-1)}+\Delta{v_{sf(m)}^n}+\Delta{v_{cor/g(m)}^n}\label{Eq.3}
\\
p_m=p_{m-1}+M_{pv(m-1/2)}(v_{m-1}^{n(m-1)}+v_m^{n(m)}\frac{T}{2})\label{Eq.4} 
\end{gather}
\vspace{-15pt}

\noindent
where $a=[\theta,\gamma,\varphi]$ represents the roll, pitch and yaw angle. The attitude matrix $C_b^n$ transforms the sampled acceleration from the carrier coordinate system (b) to the navigation coordinate system (n). $v_m^{n(m)}=[V_E,V_N,V_U]^T$ represents the east, north and up velocity at the m-th time slot in the navigation coordinate system (n). $p=[L,\lambda,h]^T$ denotes the latitude, longitude and height. And $\Delta{v_{sf(m)}^n}$ and $\Delta{v_{cor/g(m)}^n}$ in Eq.3 are the specific force velocity increment in the navigation system and the velocity increment of harmful acceleration.

\vspace{-15pt}
\begin{figure}[htbp]
\setlength{\abovecaptionskip}{-10pt}
\centerline{\includegraphics[scale=0.7]{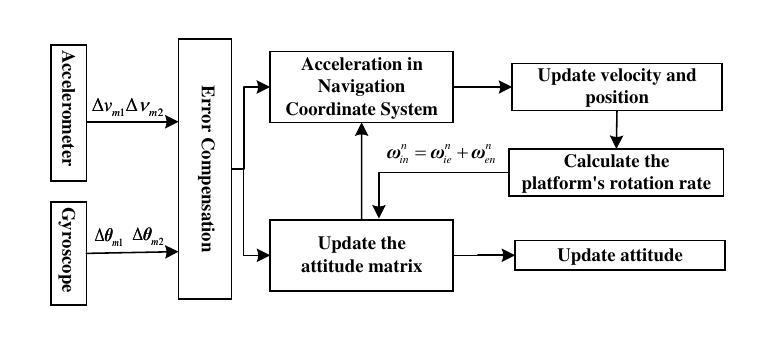}}
\caption{Basic frame diagram of SINS}
\label{fig}
\end{figure}
\vspace{-10pt}

2) Two-step WTEG Model: With the properties of the SINS and WTEG, we construct a two-step WTEG to overcome the highly dynamic topology of EUN. We use $U=\{u_1,u_2,...,u_N\}$ as the set of UAVs, where  $u_i$ represents the i-th UAV. Then the distance between two UAVs in the k-th time slot can be described as:

\vspace{-15pt}
\begin{equation}
\resizebox{0.9\hsize}{!}{$
d_{ij}(k)=\sqrt{(x_i(k)-x_j(k))^2+(y_i(k)-y_j(k))^2+(z_i(k)-z_j(k))^2} \label{Eq.4}
$}
\end{equation}
\vspace{-15pt}

We define $d_{max}$ as the maximum communication radius, then the condition to determine if a link exists can be given as:

\vspace{-20pt}
\begin{equation}
link_{ij}(k)=
\begin{cases}
1&, d_{ij}(k)\le d_{\max} \\
0&, otherwise 
\end{cases} \label{Eq.5}
\end{equation}
\vspace{-10pt}

The channel between UAVs is mainly dominated by line-of-sight (LoS) \cite{b12}, and the transmission loss is:

\vspace{-15pt}
\begin{equation}
L_{ij}(k)=(\frac{4\pi d_{ij}(k)}{\lambda})^2\xi_{LoS}=(\frac{4\pi d_{ij}(k)f}{c})^2\xi_{LoS} \label{Eq.6}
\end{equation}
\vspace{-15pt}

\noindent
where $f$ is the carrier frequency, $\lambda=c/f$ is the carrier wavelength, $c$ is the velocity of light,  and $\xi_{LoS}$ is the attenuation factors of LoS. According to the free space propagation theorem, the received signal power of the UAV can be given by:

\vspace{-15pt}
\begin{equation}
P_R^{ij}(k)=\frac{P_tG_tG_r}{L_{ij}(k)} \label{Eq.7}
\end{equation}
\vspace{-15pt}

\noindent
where $P_t$ is the transmitting power, $G_t$ is the transmitting gain, $G_r$ is the receiving gain. Then, the signal-to-noise ratio can be given by:

\vspace{-18pt}
\begin{equation}
SNR_{ij}(k)=\frac{P_R^{ij}(k)}{\sigma^2}=\frac{P_tG_tG_r}{\left(4\pi d_{ij}(k)/\lambda\right)^2\xi_{LoS}\sigma^2} \label{Eq.8}
\end{equation}
\vspace{-12pt}

\noindent
Where $\sigma^2$ is the additive white Gaussian noise. According to Shannon formula, the link capacity at the time $k$ is:

\vspace{-15pt}
\begin{equation}
R_{ij}(k)=link_{ij}(k)\cdotp B\log_2(1+SNR_{ij}(k)) \label{Eq.9}
\end{equation}
\vspace{-15pt}

Then the per-unit data transmission latency between the two UAVs is:

\vspace{-20pt}
\begin{equation}
\pi_{ij}^k=\frac{1}{R_{ij}(k)} \label{Eq.10}
\end{equation}
\vspace{-15pt}

Define the weighted matrix in the k-th time slot as:

\vspace{-15pt}
\begin{equation}
\resizebox{0.9\hsize}{!}{$
G^k=
\begin{bmatrix}
0 & \pi_{12}^k & \cdots & \pi_{1(N-1)}^k & \pi_{1N}^k \\
\pi_{21}^k & 0 & \cdots & \pi_{2(N-1)}^k & \pi_{2N}^k \\
\vdots & \vdots & \ddots & \vdots & \vdots \\
\pi_{(N-1)1}^k & \pi_{(N-1)2}^k & \cdots & 0 & \pi_{(N-1)N}^k \\
\pi_{N1}^k & \pi_{N2}^k & \cdots & \pi_{N(N-1)}^k & 0
\end{bmatrix}
$}
\end{equation}
\vspace{-10pt}

Considering that subtasks may need to be transmitted and computed across time slots, $\pi_i^{k(k+1)}$ is defined as the virtual cache delay from the k-th to the (k+1)-th time slot of node $i$:

\vspace{-10pt}
\begin{equation}
\pi_i^{k(k+1)}=\Delta{t}-t_i^k \label{Eq.12}
\end{equation}
\vspace{-15pt}

\noindent
Where $t_i^k$ is the consumed time of the i-th node in the k-th slot. The weighted matrix between two slots can be given as:

\vspace{-15pt}
\begin{equation}
\resizebox{0.9\hsize}{!}{$
G^{k(k+1)}=
\begin{bmatrix}
\pi_1^{k(k+1)} & \cdots & \infty & \cdots & \infty \\
\vdots & \ddots & \vdots & \ddots & \vdots \\
\infty & \cdots & \pi_i^{k(k+1)} & \cdots & \infty \\
\vdots & \ddots & \vdots & \ddots & \vdots \\
\infty & \cdots & \infty & \cdots & \pi_N^{k(k+1)}
\end{bmatrix}
$}
\end{equation}
\vspace{-10pt}

\vspace{-10pt}
\begin{figure}[htbp]
\setlength{\abovecaptionskip}{-5pt}
\centerline{\includegraphics[scale=0.4]{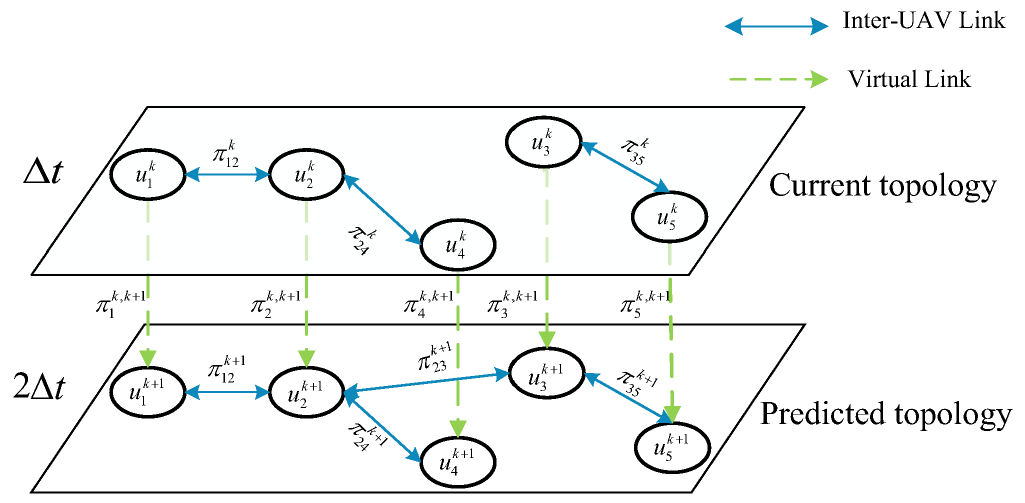}}
\caption{Two-step WTEG model}
\label{fig}
\end{figure}
\vspace{-10pt}

As Fig. 3 shows, the two-step WTEG can be represented as $\{ U,E,Graph\}$, where $\{u_1^1,u_2^1,...,u_N^1;u_1^2,u_2^2,...,u_N^2 \}$ is the set of UAV nodes in two time slots; $E$ is the edge of the two-step WTEG; $Graph$ is the matrix of the two-step WTEG, which can be given by:

\vspace{-8pt}
\begin{equation}
Graph_{k,k+1}=
\begin{bmatrix}
G^k & G^{k(k+1)} \\
\infty & G^{k+1}
\end{bmatrix} \label{Eq.14}
\end{equation}
\vspace{-15pt}

\subsection{Collaborative Computing Strategy based on Task-Driven}\label{B}

The task $\varPhi$ can be characterized as a DAG as shown in Fig. 4, and the task scheduling problem is transformed into a task DAG to WTEG mapping problem. 

In general, the task $\Phi$ can be represented as a DAG, $\Phi=\{ \Omega,\Gamma\}$, where $\Omega=\{\omega_1,...,\omega_l \}$ is the set of nodes,  $\Gamma$ is the set of edges. The subtasks are represented by the nodes of the DAG, while the edges indicate the execution order among the subtasks. In $\Phi=\{ \Omega,\Gamma\}$, $\omega_1$ is the single starting subtask, $\omega_l$ is the terminal subtask, and $\omega_i$ are the other subtasks. Besides, subtask  $\omega_i \in \Omega$ is represented by the tuple $\{D_i,C_i,\xi_i \}$, where $D_i$ represents the input data size, $\xi_i \in (0,1]$ represents data scaling factor, $C_i$ represents the required amount CPU cycles to complete the subtask. Moreover, we define $\delta_i=(C_i/D_i)$cycle/bit as the computation complexity of subtask \cite{b13}; $\Theta_{\uparrow}(\omega_{i})=\left\{\omega_{j}\mid(\omega_{j},\omega_{i})\in\Gamma\right\}$ as the set of predecessor subtasks of $\omega_i \in \Omega$. Thus, $D_i$ can be given by:

\vspace{-15pt}
\begin{equation}
D_i=\sum_{\omega_j\in\Theta_\uparrow(\omega_i)}D_j\xi_j \label{Eq.15}
\end{equation}
\vspace{-12pt}

1) $\Phi$ to $U$ Mapping Rules:We use $B: \Omega \to U$ to represent the mapping from subtasks $\Omega$ to UAV nodes $U$. Then B should meet the requirements shown in Eq.17. Map the starting subtask $\omega_1$ to the task initiator $u_1$; map the terminal subtask $\omega_l$ to the result receiver $u_N$; map the intermediate subtasks to any other UAVs. The execution of subtasks may need to cross time slots and the task data will be cached to the next time slot through the virtual link. Therefore, for any subtask $\omega_i \in \Omega$, it will be mapped to the $\chi_i+1$ replica nodes, where $\chi_i$ is the number of cross-time slots.

\vspace{-15pt}
\begin{equation}
\resizebox{1.0\hsize}{!}{$
B(\omega_i)=
\begin{cases}
\left\{u_1^1,\cdots u_1^{1+\rho_i}\mid1+\chi_i\leq2\right\}, & \text{if } \omega_i=\omega_1\\
\left\{u_N^{k},\cdots u_N^{k+\rho_i}\mid k\geq1,k+\chi_i\leq2\right\}, & \text{if } \omega_i=\omega_l\\
\left\{u_d^{k},\cdots u_d^{k+\rho_i}\mid k\geq1,k+\chi_i\leq2,1\leq d\leq N\right\}, & \text{otherwise}
\end{cases}
$}
\end{equation}
\vspace{-10pt}

2) $\Gamma$ to $E$ Mapping Rules: We define $Z: \Gamma \to E$ as the mapping from $\Gamma$ to $E$. For any directed edge $(\omega_i,\omega_j) \in \Gamma$, Z maps it to the shortest path $Path_{end(B(\omega_{i})),start(B(\omega_{j}))}$ between the last node $end(B(\omega_i))$ of the node set $B(\omega_i)$ to the first node $start(B(\omega_j))$ of the node set $B(\omega_j)$.

\vspace{-10pt}
\begin{equation}
Z(\omega_i,\omega_j)=Path_{end(B(\omega_i)),start(B(\omega_j))} \label{Eq.17}
\end{equation}
\vspace{-15pt}

3)Optimization Problem Formulation: Based on the aftermentioned mapping rules, the delay model of UAV cluster collaborative computing is given according to the mapping results. The processing delay of subtask $\omega_i$ can be given by:

\vspace{-15pt}
\begin{equation}
T(\omega_i)=T_{comp}(\omega_i)+T_{accu}(\omega_i) \label{Eq.18}
\end{equation}
\vspace{-15pt}

\noindent
where $T_{comp}(\omega_i)$ is the computing latency of subtask $\omega_i$ as shown in Eq. 20; $\rho_{B(\omega_i)}$ is the computing capacity of $B(\omega_i)$ to indicate the cycle of CPU per second; $T_{accu}(\omega_i)$ is the accumulated latency as shown in Eq. 21; $d_{end(B(\omega_j)),start(B(\omega_i))}$ is the transmission delay of $\forall\omega_j\in\phi_\uparrow(\omega_i)$ transmitting unit bit to $\omega_i$ along the shortest path $Path_{end(B(\omega_{i}))start(B(\omega_{j}))}$.

\vspace{-8pt}
\begin{equation}
T_{comp}(\omega_i)=\frac{D_i\delta_i\xi_i}{\rho_{B(\omega_i)}} \label{Eq.19}
\end{equation}
\vspace{-10pt}

\vspace{-15pt}
\begin{equation}
T_{accu}(\omega_i)=\max_{\omega_j\in\phi_\uparrow(\omega_i)}[T(\omega_j)+d_{end(B(\omega_j)),start(B(\omega_i))}D_j\xi_j]\label{Eq.20}
\end{equation}
\vspace{-15pt}

Therefore, the total processing latency of task $\Phi$ can be given by: 

\vspace{-15pt}
\begin{equation}
\resizebox{1.0\hsize}{!}{$
T(\Phi)=T(\omega_1)=\max_{w_j\in\phi_1(w_l)}[T(w_j)+d_{end(B(w_j)),start(B(w_l))}D_j\xi_j]+\frac{D_l\delta_l\xi_l}{\rho_{B(w_l)}}
$}
\end{equation}
\vspace{-10pt}

For the same task $\Phi$ and network topology, there are many different mapping results that satisfy the mapping rules, which lead to different processing delays. Therefore, we need to find the mapping results with the lowest processing delay. Any mapping results can be uniquely represented as a decision matrix $X$ with $l$ rows $N\times2$ columns, which can be given by:

\vspace{-20pt}
\begin{equation}
\resizebox{0.9\hsize}{!}{$
X=
\begin{bmatrix}
x(\omega_1,u_1^1) & \cdots & x(\omega_1,u_N^1) & \cdots & x(\omega_1,u_N^2) \\
x(\omega_2,u_1^1) & \cdots & x(\omega_2,u_N^1) & \cdots & x(\omega_2,u_N^2) \\
\vdots & \ddots & \vdots & \ddots & \vdots \\
x(\omega_l,u_1^1) & \cdots & x(\omega_l,u_N^1) & \cdots & x(\omega_l,u_N^2)
\end{bmatrix}\label{Eq.22}
$}
\end{equation}
\vspace{-8pt}

\noindent
Where $x(\omega_i,u_d^k)\in X$ represents whether $\omega_i$ is mapped to the $u_d^k$. If $x(\omega_{i},u_{d}^{k})=1$, $\omega_i$ is mapped to the $u_d^k$, or $\omega_i$ is not mapped to the $u_d^k$.

\vspace{-12pt}
\begin{equation}
x(\omega_i,u_d^k)\in
\begin{Bmatrix}
0,1
\end{Bmatrix},\forall\omega_i\in\Omega,\forall u_d^k\in U,k\in\{1,2\} \label{Eq.23}
\end{equation}
\vspace{-12pt}

For any subtask $\omega_i \in \Omega$, it will be mapped to $\rho_i+1$ replica nodes of the same UAV from Eq. 17.

\vspace{-15pt}
\begin{equation}
\sum_{u_d^k\in U}x(\omega_i,u_d^k)=\chi_i+1,\forall\omega_i\in\Omega,\chi_i\in\{0,1\} \label{Eq.24}
\end{equation}
\vspace{-8pt}

Therefore, the total processing delay of task $\Phi$ can be given by:

\vspace{-15pt}
\begin{equation}
T(X)=T_{accu}(\omega_l)+\sum_{u_d^2\in U}\frac{D_i\delta_i\xi_i}{\rho_{u_d^2}}x(\omega_l,u_d^2) \label{Eq.25}
\end{equation}
\vspace{-10pt}

According to the analysis above, the delay optimization problem can be represented as:

\vspace{-10pt}
\begin{equation}
\begin{aligned}
 & X=\arg\min(T(X)) \\
 & s.t.:(24)(25)
\end{aligned}\label{Eq.26}
\end{equation}

\section{\bfseries Binary particle swarm optimization algorithm}

In order to solve the NP-hard problem in Eq.27, BPSO algorithm is adopted, which is a swarm intelligent search algorithm mainly used to optimize the constraint problems in discrete space. Its particle swarm search area changes with the increase of time slot, making it suitable for the dynamic UAV network environment.

We denote the total particle as $M$, and $I$ represents the maximum iterations of the algorithm. The position and the flying speed of the m-th particle at the i-th iteration can be given as: 

\vspace{-15pt}
\begin{equation}
\resizebox{0.85\hsize}{!}{$
X_{m}^{i}=(x_{m}^{i}(\omega_{1},u_{1}^{1}),\cdots,x_{m}^{i}(\omega_{l},u_{N}^{1}),\cdots,x_{m}^{i}(\omega_{l},u_{N}^{2}))
$}
\end{equation}
\vspace{-10pt}

\vspace{-20pt}
\begin{equation}
\resizebox{0.85\hsize}{!}{$
V_{m}^{i}=(v_{m}^{i}(\omega_{1},u_{1}^{1}),\cdots,v_{m}^{i}(\omega_{l},u_{N}^{1}),\cdots,v_{m}^{i}(\omega_{l},u_{N}^{2}))
$}
\end{equation}
\vspace{-15pt}

Besides, the flying speed of the m-th($m \leq M$) particle at the i-th($i \leq I$) iteration is updated by:

\vspace{-15pt}
\begin{equation}
V_m^{i+1}=\mu V_m^i+\alpha_1\beta_1(p_{M_{best}}-X_m^i)+\alpha_2\beta_2(g_{best}-X_m^i)\label{Eq.29}
\end{equation}
\vspace{-15pt}

\noindent
Where $\mu$ is the inertia weight, $\alpha_1$ and $\alpha_2$ are learning factors, $\beta_1$ and $\beta_1$ are random values distributed in the interval [0,1],  $P_{M_{best}}$ and $g_{best}$ are the local optimal position and the global optimal position of the m-th particle.

Use sigmoid function to map particle velocity to [0,1] interval, which can be given by:

\vspace{-12pt}
\begin{equation}
S(V_m^{i+1})=\frac{1}{1+\exp(-V_m^{i+1})}\label{Eq.30}
\end{equation}
\vspace{-10pt}

Therefore, the flying position of the $m \leq M$th particle is updated by:

\vspace{-16pt}
\begin{equation}
X_m^{i+1} =
\begin{cases}
1, & \text{if } S(V_m^{i+1}) \geq \text{rand()} \\
0, & \text{otherwise}
\end{cases}\label{Eq.31}
\end{equation}
\vspace{-12pt}

\noindent
Where $S(V_m^{i+1})$ is the probability of $X_m^{i+1}=1$. $rand()$ is a number randomly generated in the interval $[0,1]$. Finally,the fitness value is calculated by:

\vspace{-10pt}
\begin{equation}
f(X_m^{i+1})=T(X_m^{i+1})\label{Eq.32}
\end{equation}
\vspace{-15pt}

\section{\bfseries SIMULATION RESULTS}

To validate the effectiveness of the proposed scheme, a series of experiments are conducted. The simulation parameters are configured as follows: each time slot is set to 4 seconds, and the number of UAVs is 9. For the algorithm, the particle count ($M$) is 100, and the maximum number of iterations ($I$) is 100, the inertia weight $\mu$ is 1.5, and both learning factors ($\alpha_1$ and $\alpha_2$) are set to 1. According to the literature [14][15][16][17], the parameters are summarized in TABLE 1. Additionally, two types of task DAG with different scales are shown in Fig. 4.

\vspace{-10pt}
\begin{table}[htbp]
\setlength{\abovecaptionskip}{-3pt}
\caption{System Parameters}
\label{基本参数} 
\centering  
\begin{tabular}{|c|c|c|c|}
\hline
Parameter & Value & Parameter & Value \\
\hline
Pt & 0.05W & Gt & 3dB \\
\hline
Gr & 3dB & f & 2.4GHz \\
\hline
B & 20MHz & $d_{max}$ & 6000m \\
\hline
$\sigma^2$ & -100dBm & $\rho_{B(\omega_i)}$ & [500,1200]MHz\\
\hline
$\delta_{i}$ & 1900/8cycle/bit & $\xi_i$  & 0.8 \\
\hline
\end{tabular}
\end{table}
\vspace{-18pt}

\begin{figure}[htbp]
\setlength{\abovecaptionskip}{-10pt}
\centerline{\includegraphics[scale=0.32]{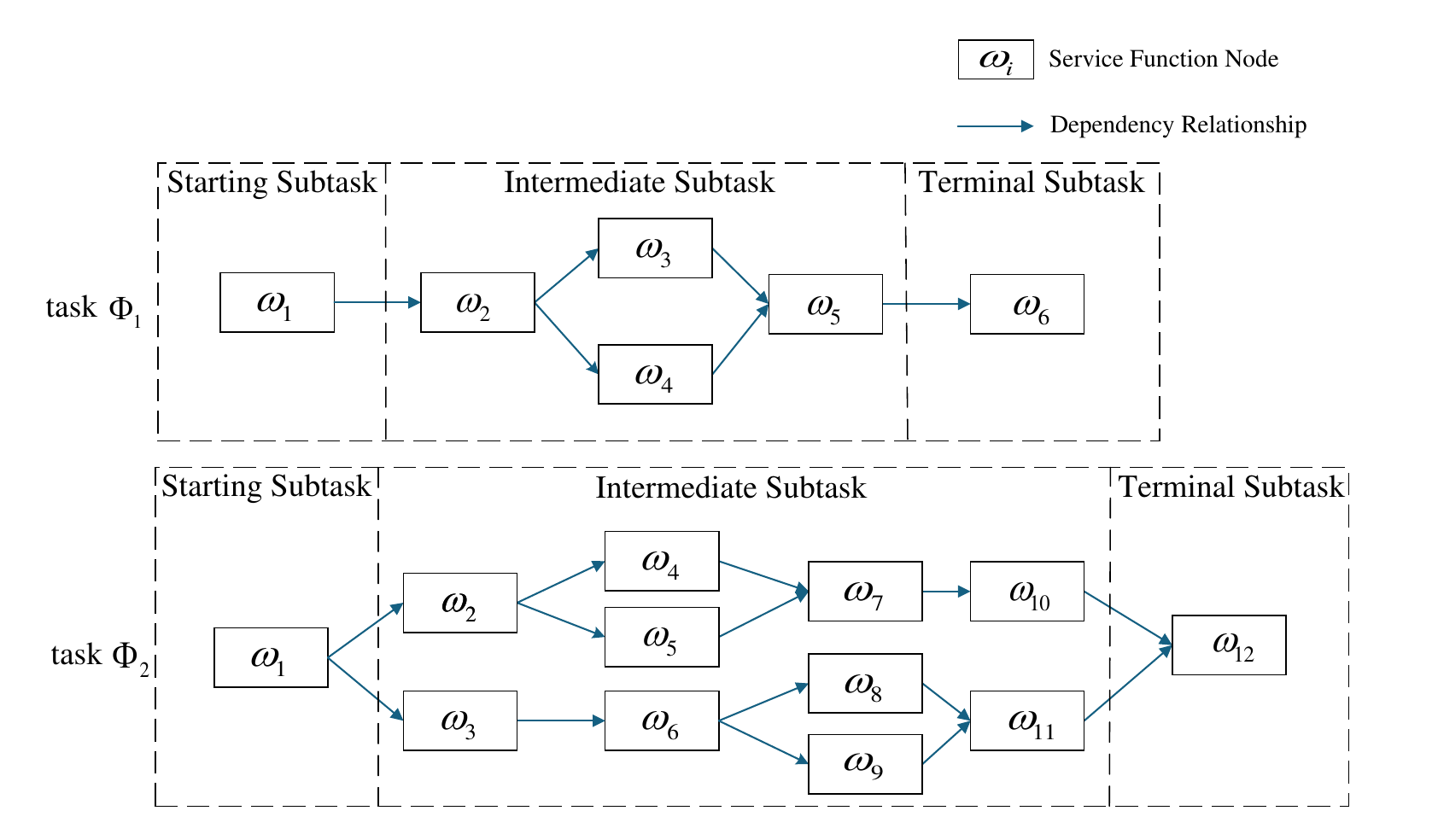}}
\caption{The task DAG models with different scales}
\label{fig}
\end{figure}
\vspace{-10pt}

\subsection{Position Prediction Accuracy of SINS}\label{AA}

\vspace{-15pt}
\begin{figure}[htbp]
\setlength{\abovecaptionskip}{-5pt}
\centerline{\includegraphics[scale=0.6]{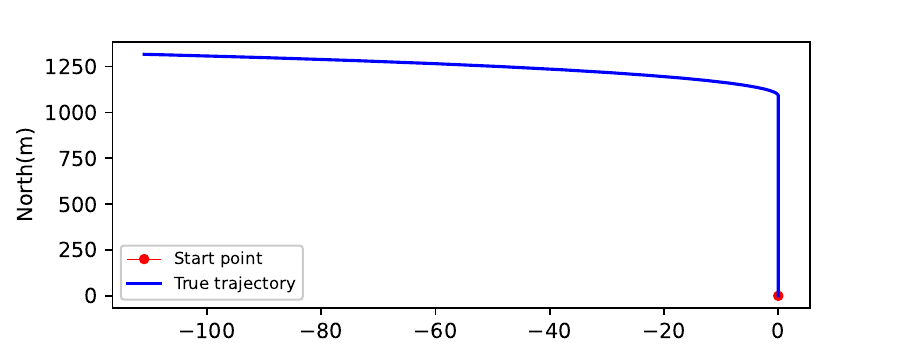}}
\caption{The real trajectory of UAV}
\label{fig}
\end{figure}
\vspace{-10pt}

The sensor sampling interval in SINS is set to 0.1s, with a total duration of 240s. The actual trajectory of a single UAV is shown in Fig. 5, where the latitude and longitude of the initial position is $(29^{\circ},106^{\circ})$, maintaining a constant height of 450m. Over the entire process, the UAV travels 1316m northward and 110m westward.

Fig. 6 presents the prediction error of the SINS model relative to the actual AVP (Attitude, Velocity, and Position) data. The simulation results indicate that the attitude error remains below $0.1^{\circ}$, the velocity error is within 1m/s, and the position error is less than 20m within the first 100s. However, prediction errors accumulate over time, with the north-direction position error and altitude error reaching approximately 7.6$\%$ and 11$\%$ of the total actual distance respectively by 240s. Therefore, the position prediction of UAV based on SINS exhibits high accuracy over short durations, which is suitable for dynamic network operations with rapid data transmission.

\vspace{-15pt}
\begin{figure}[htbp]
\setlength{\abovecaptionskip}{-2pt}
\centerline{\includegraphics[scale=0.4]{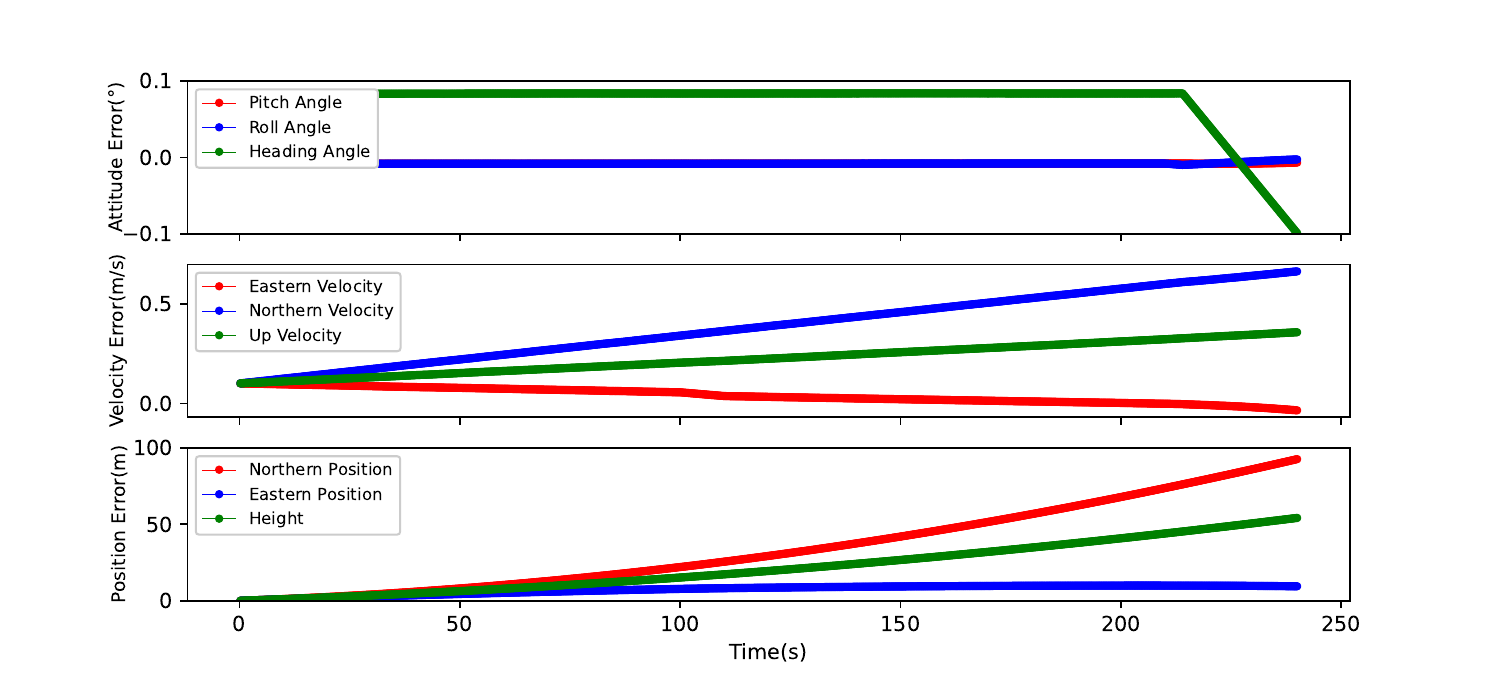}}
\caption{Difference from Actual AVP data}
\label{fig}
\end{figure}
\vspace{-15pt}

\subsection{Latency Performance of Cooperative Computing for UAVs}\label{BB}

Fig. 7 illustrates the latency performance comparison of three computation architectures (i.e., cloud computing, local computing, and collaborative computing) for two different task scales, with the input data size ranging from 0 Mb to 5 Mb.

\vspace{-10pt}
\begin{figure}[htbp]
\setlength{\abovecaptionskip}{-5pt}
\centerline{\includegraphics[scale=0.45]{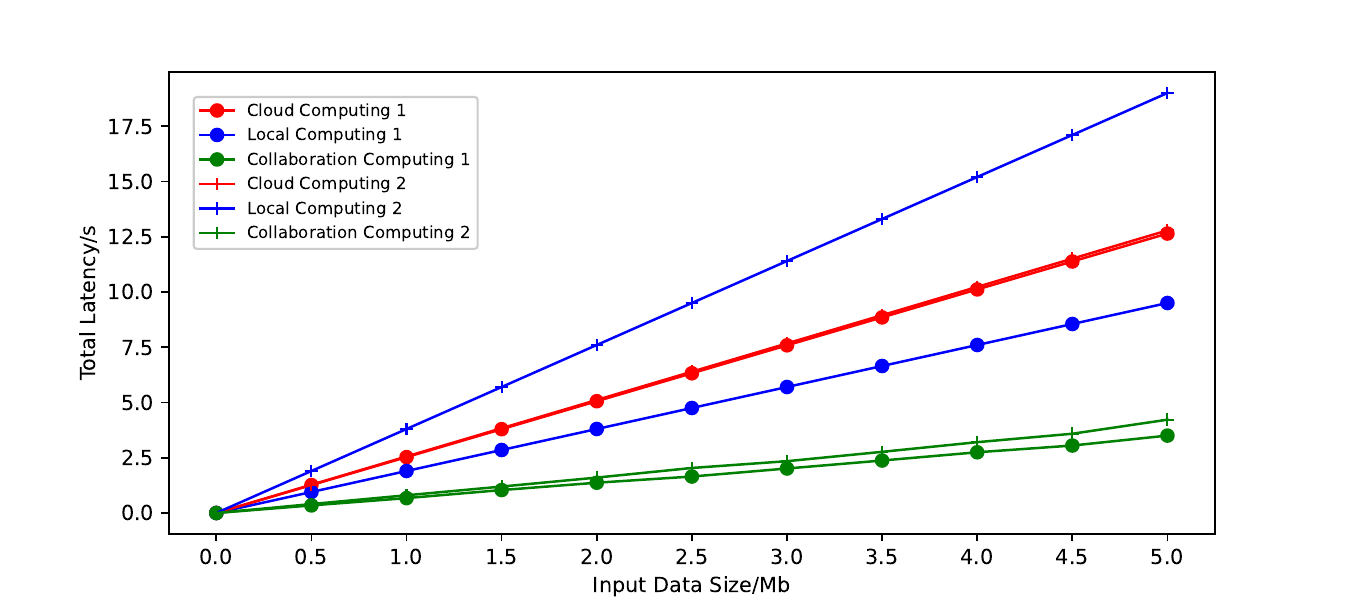}}
\caption{Latency performance comparision of three computation architectures}
\label{fig}
\end{figure}
\vspace{-5pt}

With the increase of the input data size, the cloud computing curve is significantly higher than the local computing method for task $\Phi_1$ but lower for task $\Phi_2$. This is because task $\Phi_2$ involves more subtasks, thus requiring greater computing resources. Cloud computing relies on long-distance cloud servers with high computing ability, resulting in large transmission latency but lower computation latency. Furthermore, the local computing strategy is constrained by the limited computing ability of a single UAV. The latency of collaborative computing consistently achieves the lowest latency for both tasks. This superior performance is attributed to lower transmission latency within compact drone clusters and reduced computation latency depending on the collective computational power of numerous drones. When the input data size is 5 Mb, collaborative computing shows a latency improvement of 72.84$\%$, 63.85$\%$ over cloud computing and 66.98$\%$, 77.78$\%$ over local computing for task$\Phi_1$, task$\Phi_2$ respectively. Therefore, collaborative computing is suitable for latency-sensitive tasks in disaster scenes.

\subsection{Relationship between Task Processing Delay and Computation Complexity}\label{CC}

Take the task $\Phi_1$ shown in Fig.4 as an example, Fig. 8 shows the change trend of task processing latency , where the computation complexity ranges from $0.2\delta_i$ to $1.8\delta_i$ and the input data size ranges from 0 Mb to 5Mb.

\vspace{-10pt}
\begin{figure}[htbp]
\setlength{\abovecaptionskip}{-5pt}
\centerline{\includegraphics[scale=0.3]{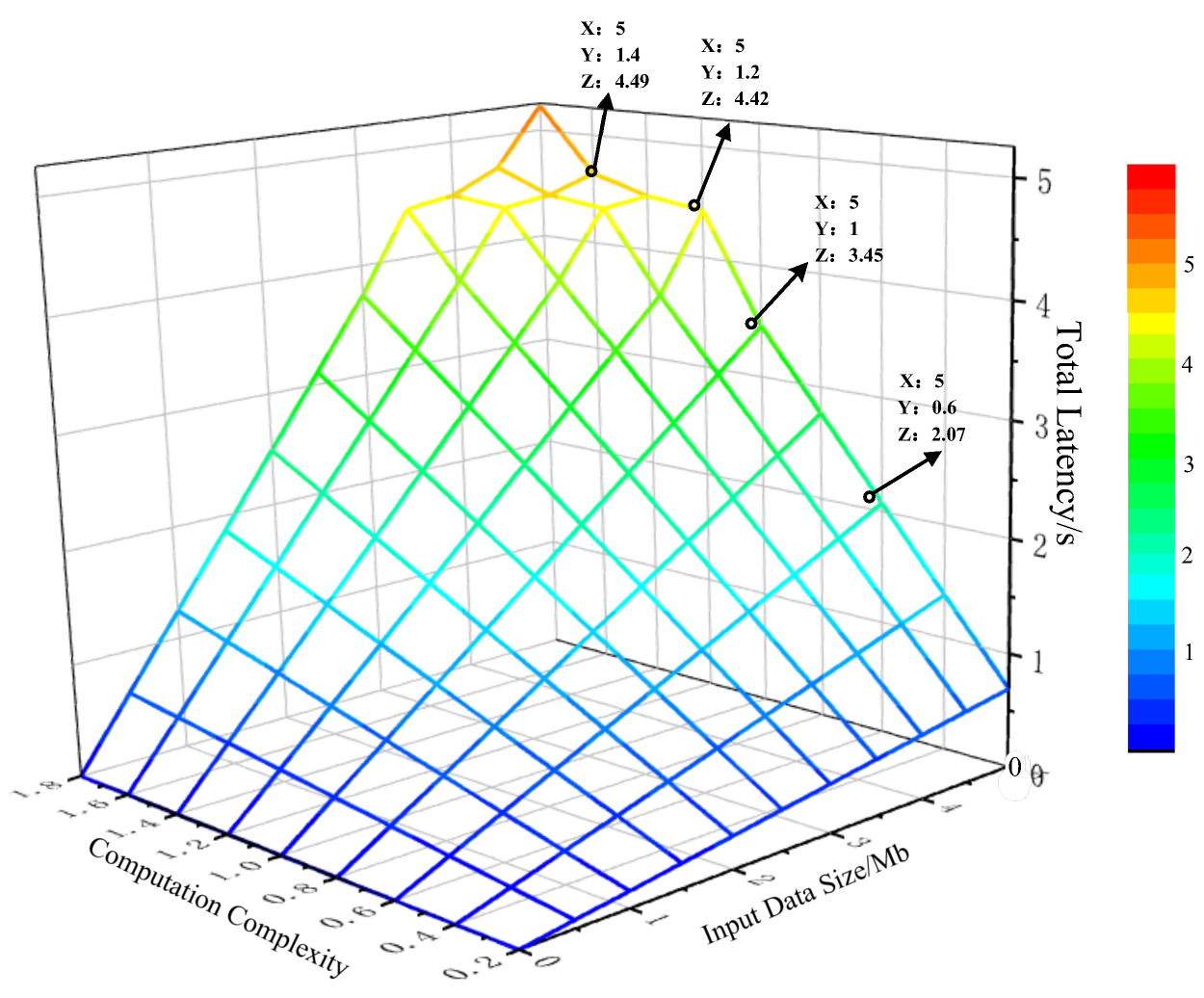}}
\caption{Latency performance under different computation complexity}
\label{fig}
\end{figure}
\vspace{-5pt}

We can observe that for a given input data size, the task processing delay increases with the computation complexity, and the increasing trend varies with data volume constraints. When the input data size is 1Mb, the task latency slightly increases from 0.07s to 0.62s as the computation complexity increases from $0.2\delta_i$ to $1.8\delta_i$. This is because the task computation latency gradually increases with the computation complexity. However, when the input data size reaches 5Mb, the task latency jumps from 0.69s to 5.22s, especially when the computation complexity increases from $\delta_i$ to $1.2\delta_i$, the total task processing delay jumps from 3.45s to 4.42s. The reason is that the task cannot be completed in the first time slot and the current UAV needs to cache data until the next time slot, which leads to a significant increase in the transmission delay.

\subsection{Latency Performance of Different Algorithms}\label{DD}

Fig. 9 shows the latency performance of the BPSO algorithm compared with typical load balancing algorithms (i.e.,Weighted Round Robin(WRR), Greedy Load Balance(Greedy-LB), Pick-KX) under two kinds of task scales. 

\begin{figure}[h]
    \centering
    \subfigure[Task $\varPhi_1$]
    {\begin{minipage}{0.49\linewidth}
        \centering
        \includegraphics[scale=0.32]{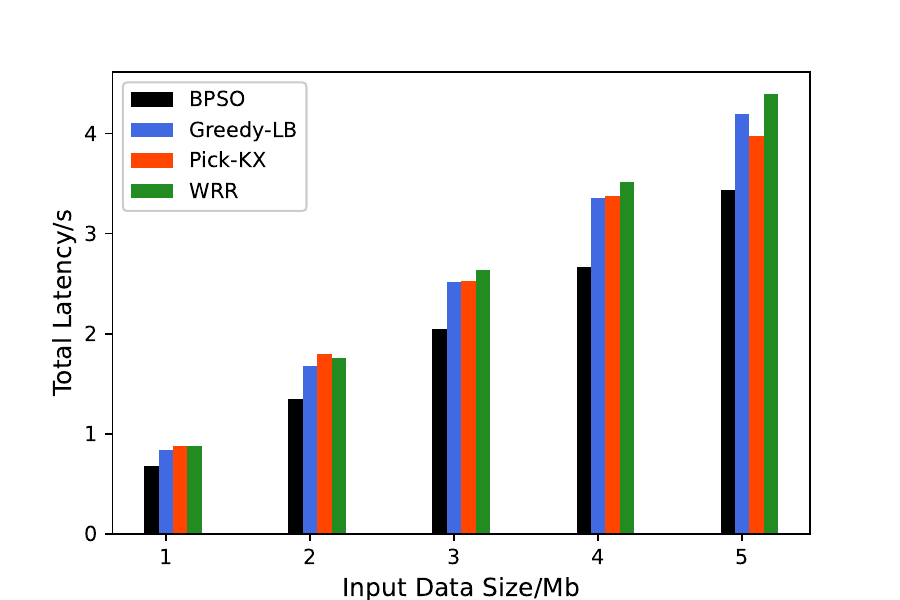}
        \label{图片label}
    \end{minipage}}
    \hfill
    \subfigure[Task $\varPhi_2$]
    {\begin{minipage}{0.49\linewidth}
        \centering
        \includegraphics[scale=0.32]{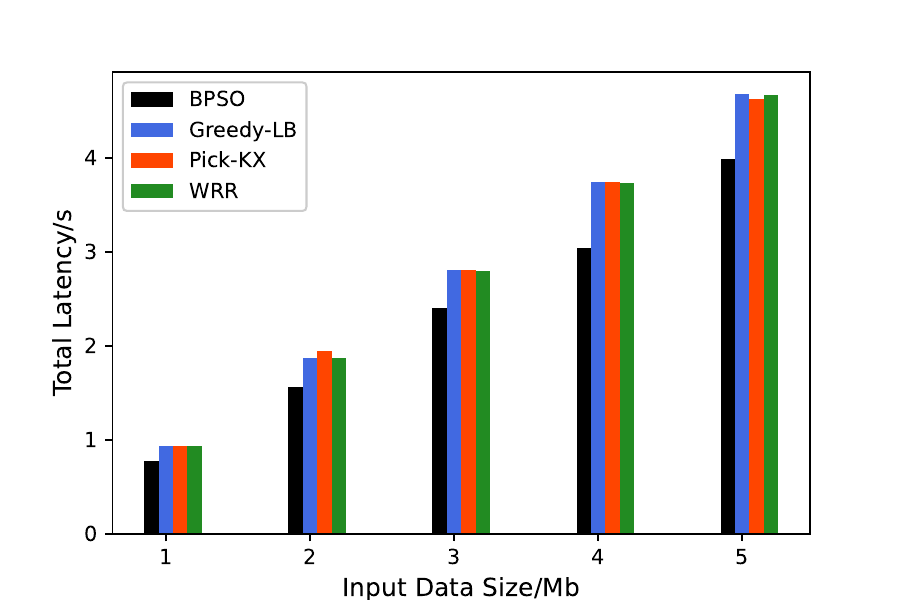}
        \label{图片label}
    \end{minipage}}
    \caption{Latency performance under different algorithms}
\end{figure}

It is evident that the BPSO algorithm is superior to the typical load balancing for different task scales. This is because the WRR algorithm only considers the computation capabilities of the drones, while the Greedy-LB algorithm focuses solely on the current load conditions. The two algorithms both ignore the impact of transmission link capabilities. Besides, the pick-KX algorithm randomly distributes subtasks to drones without taking the computing power into account. In contrast, the BPSO algorithm jointly considers computation and communication capabilities, thereby achieving reasonably high latency performance. For task $\Phi_1$ with a 5 Mb data size, results suggest that the latency performance of the BPSO algorithm improved by 8.36$\%$, 22.04$\%$, 27.96$\%$ compared with the Greedy-LB, WRR, and Pick-KX algorithms respectively.

\subsection{Impact of SINS Prediction on Task Success Rate}\label{EE}

Fig. 10 illustrates the task success rate of One-Step Prediction based SINS against a baseline without prediction, where the input data size varied from 1 Mb to 10 Mb, and 200 task requests were sent at each instance. The task success rate is defined as the percentage of tasks successfully transmitting data to the result receiver out of all initiated task requests. 

\vspace{-5pt}
\begin{figure}[htbp]
\setlength{\abovecaptionskip}{-3pt}
\centerline{\includegraphics[scale=0.5]{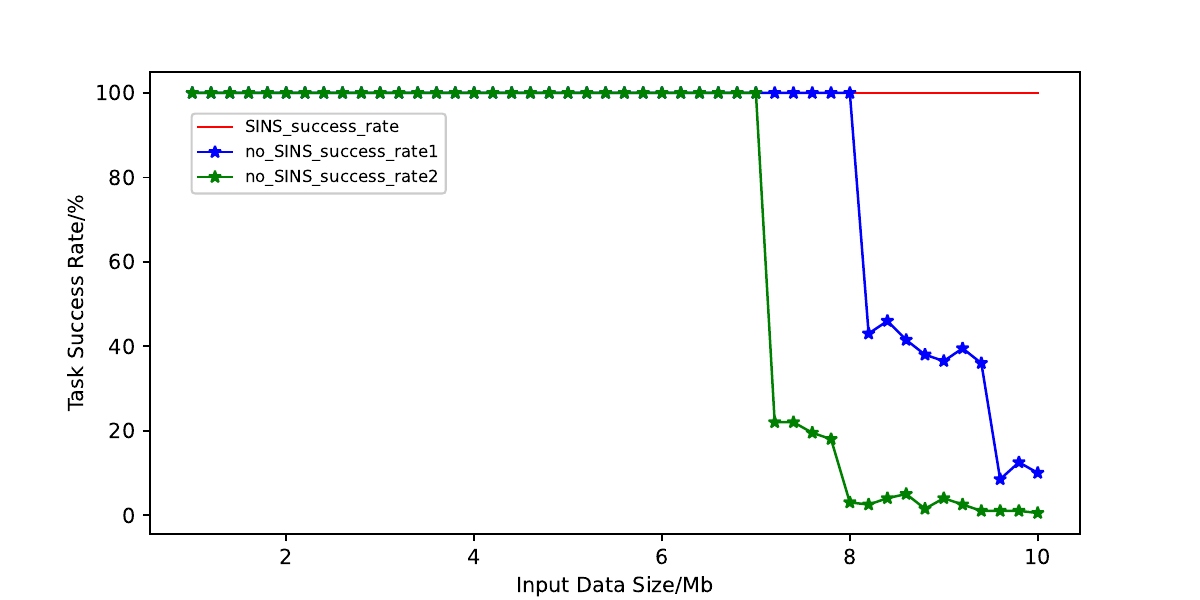}}
\caption{Task success rate based SINS and no-SINS}
\label{fig}
\end{figure}
\vspace{-5pt}

The results indicate that the task success rate using SINS always remains at 100$\%$. The reason is that SINS accurately predicts the future positions of the drones, ensuring that the selected drones can operate normally in the second time slot to offload service in advance. In contrast, the strategy without SINS experiences a sharp decline in success rate when the data size exceeds a certain threshold (8 Mb for task $\Phi_1$ and 7.2 Mb for task $\Phi_2$). Furthermore, as the number of cross-slot subtasks increases, the service success rate plummets. This decline is primarily because the link connection relationship in the second time slot is uncertain, and the success rate of subtask is $(\frac{1}{2})^n$ where n is the number of predecessor nodes. When the input data size is 10Mb, the task success rate using SINS improves by more than 90$\%$ compared with no-SINS. In summary, One-Step Prediction based SINS effectively mitigates service failures caused by dynamic UAV networks.

\section{\bfseries CONCLUSION}

To address the problem of real-time variation of UAVs' flight trajectories in emergency rescue scenarios, SINS is employed for one-step prediction of drone positions. Building upon this prediction, two-step WTEG model is constructed to overcome the dynamic changes in network topology. Furthermore, a DAG-WTEG task mapping model is established, and the BPSO algorithm is used to select the optimal delay service mapping strategy, which enables drones to compute while transmitting. Simulation results indicate that for different task scales the task success rate under one-step prediction is significantly improved, and the proposed collaborative computing strategy exhibits excellent delay performance.

\end{document}